\documentclass[prl,floatfix,showpacs,twocolumn,tightenlines,nofootinbib,superscriptaddress]{revtex4-1}

\usepackage{ulem}

\usepackage[utf8]{inputenc}
\usepackage[T1]{fontenc}
\usepackage{braket}
\usepackage{graphicx}
\usepackage{booktabs}
\usepackage{mathtools}
\usepackage{amssymb}
\usepackage{amsmath}
\usepackage{amsfonts}
\usepackage{mathrsfs}
\usepackage{color}
\usepackage{amsthm}
\usepackage{psfrag}
\usepackage{ifsym}
\usepackage{dsfont}
\usepackage{enumerate}
\usepackage{enumitem}
\usepackage{multirow} 
\usepackage{hyperref}
\usepackage{lipsum}
\usepackage{float}
\usepackage[separate-uncertainty=true]{siunitx}
\usepackage{setspace}
\hypersetup{colorlinks=true,linkcolor=red,citecolor=blue}
\usepackage[sort&compress]{natbib}
\usepackage[dvipsnames]{xcolor}
\usepackage{comment}

\newcommand*\hgzero{\includegraphics[height=7pt]{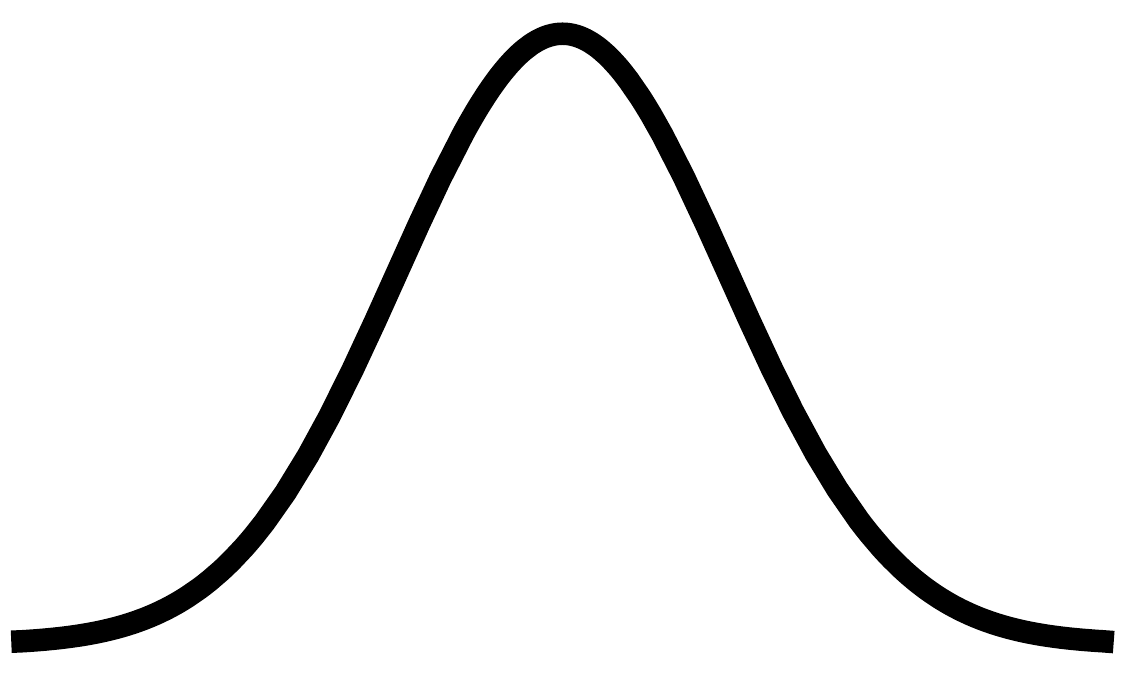}}
\newcommand*\hgone{\includegraphics[height=7pt]{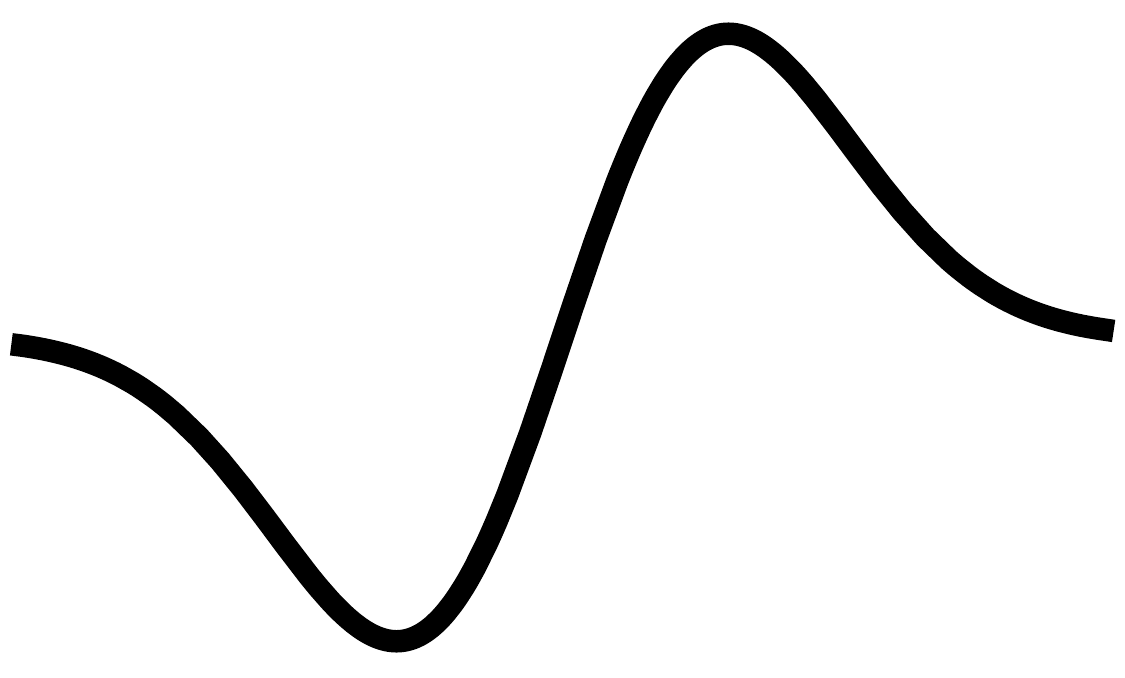}}

\begin{document}
\title{Hyperentanglement in structured quantum light}

\author{Francesco Graffitti}
\email[Corresponding author: ]{fraccalo@gmail.com}
\affiliation{Institute of Photonics and Quantum Sciences, School of Engineering and Physical Sciences, Heriot-Watt University, Edinburgh EH14 4AS, United Kingdom}

\author{Vincenzo D'Ambrosio}
\email[Corresponding author: ]{vincenzo.dambrosio@unina.it}
\affiliation{Dipartimento di Fisica, Università di Napoli Federico II, Complesso Universitario di Monte S. Angelo, Via Cintia, 80126 Napoli, Italy}

\author{Massimiliano Proietti}
\affiliation{Institute of Photonics and Quantum Sciences, School of Engineering and Physical Sciences, Heriot-Watt University, Edinburgh EH14 4AS, United Kingdom}

\author{Joseph Ho}
\affiliation{Institute of Photonics and Quantum Sciences, School of Engineering and Physical Sciences, Heriot-Watt University, Edinburgh EH14 4AS, United Kingdom}

\author{Bruno Piccirillo}
\affiliation{Dipartimento di Fisica, Università di Napoli Federico II, Complesso Universitario di Monte S. Angelo, Via Cintia, 80126 Napoli, Italy}

\author{Corrado de Lisio}
\affiliation{Dipartimento di Fisica, Università di Napoli Federico II, Complesso Universitario di Monte S. Angelo, Via Cintia, 80126 Napoli, Italy}
\affiliation{CNR-SPIN U.O.S. di Napoli, Via Cintia 2, 80126 Napoli, Italy}

\author{Lorenzo Marrucci}
\affiliation{Dipartimento di Fisica, Università di Napoli Federico II, Complesso Universitario di Monte S. Angelo, Via Cintia, 80126 Napoli, Italy}

\author{Alessandro Fedrizzi}
\affiliation{Institute of Photonics and Quantum Sciences, School of Engineering and Physical Sciences, Heriot-Watt University, Edinburgh EH14 4AS, United Kingdom}

\begin{abstract}
Entanglement in high-dimensional quantum systems, where one or more degrees of freedom of light are involved, offers increased information capacities and enables new quantum protocols.
Here, we demonstrate a functional source of high-dimensional, noise-resilient hyperentangled states encoded in time-frequency and vector-vortex structured modes, which in turn carry single-particle entanglement between polarisation and orbital angular momentum.
Pairing nonlinearity-engineered parametric downconversion in an interferometric scheme with spin-to-orbital-angular-momentum conversion, we generate highly entangled photon pairs at telecom wavelength that we characterise via two-photon interference and quantum state tomography, achieving near-unity visibilities and fidelities.
While hyperentanglement has been demonstrated before in photonic qubits, this is the first instance of such a rich entanglement structure involving spectrally and spatially structured light, where three different forms of entanglement coexist in the same biphoton state.
\end{abstract}

\maketitle

Photonic platforms are a natural choice for many quantum applications
owing to their advantages as low-noise quantum systems with high-fidelity control and suitability for long-distance transmission.
Binary encoding has been the most common choice from the early, proof-of-principle experiments to the most recent photonic quantum protocols. 
However, there are scenarios that benefit from expanding the system dimensionality, e.g. for enhancing information capacity, noise resilience and robustness against external attacks in quantum cryptography~\cite{Erhard2018,Cozzolino19}.
Intrinsically high-dimensional degrees of freedom (DOF) of light---such as orbital angular momentum (OAM), time and frequency---enable a larger quantum alphabet in a single photon state.
The combination of two or more DOFs of light---including entanglement across them, namely \textit{hyperentanglement}~\cite{PhysRevLett.95.260501}---allows further expansion of the Hilbert space while providing easy access to the individual subsystems for selective control and measurements, improving existing protocols or enabling new ones~\cite{D'Ambrosio2012,PhysRevX.5.041017}.
Hyperentanglement in particular enables protocols like complete Bell-state analysis \cite{PhysRevA.68.042313,PhysRevLett.96.190501,PhysRevA.75.042317} and logic gates simplification~\cite{Lanyon2009}, and has been used in cluster state generation~\cite{PhysRevA.81.052301,Ciampini2016} as well as in testing quantum foundations~\cite{PhysRevLett.97.140407}. 
Moreover, hyperentangled systems have been successfully used for demonstrations of quantum dense coding \cite{Barreiro2008} and teleportation of multiple DOFs of a single photon~\cite{Wang2015}.

Photonic hyperentangled states have been demonstrated in different encoding regimes, as polarisation, time and frequency bins, path, and OAM~\cite{PhysRevLett.95.260501,PhysRevA.75.042317,Xie2015}. 
While entanglement between three DOFs have been achieved in the past, e.g. for significantly expanding the accessible Hilbert-space dimension in two- or multi-photon experiments \cite{PhysRevLett.95.260501,PhysRevLett.120.260502},
the generation of hyperentanglement of spatially or spectrally structured light has, to our knowledge, so far remained elusive, probably due to the complexity in accessing such encoding regimes.
However, structured light modes---where one or more degrees of freedom are modulated into custom light fields---are a useful resource in quantum photonics applications, spanning communication, metrology and imaging among others~\cite{Rubinsztein_Dunlop_2016}.
In this work, we fill this gap combining a nonlinearity engineering technique~\cite{Graffitti_2017,PhysRevLett.124.053603} with a spin-to-orbital-angular-momentum conversion scheme~\cite{PhysRevLett.96.163905} to generate and characterise a biphoton state that exhibits complex entanglement between spectrally and spatially structured light.
We produce hyperentanglement between time-frequency modes (TFM)---temporal/spectral envelopes of the electric field of the photons~\cite{PhysRevX.5.041017,PhysRevLett.124.053603}, and vector vortex beams (VVB)---spatially structured beams characterised by a non-uniform polarisation pattern on their transverse profile~\cite{DENNIS2009293,Cardano:12,PhysRevA.100.063842}. 
Due to their resilience to different noise sources, both TFMs~\cite{Eckstein:11,ding2019highdimensional} and VVBs~\cite{D'Ambrosio2012,Farias2015} provide ideal encodings in free-space communication schemes, while sources that generate polarisation-TFM hyperentanglement could be immediately deployed in current telecom networks, being both degrees of freedom already compatible with fibre propagation.
It has recently been shown that VVB can be supported in particular fibres~\cite{10.1117/1.AP.1.4.046005}, opening the prospect of fibre-based communication schemes exploiting the full capabilities of our scheme.

\begin{figure}[t!]
\begin{center}
\includegraphics[width=0.95\columnwidth]{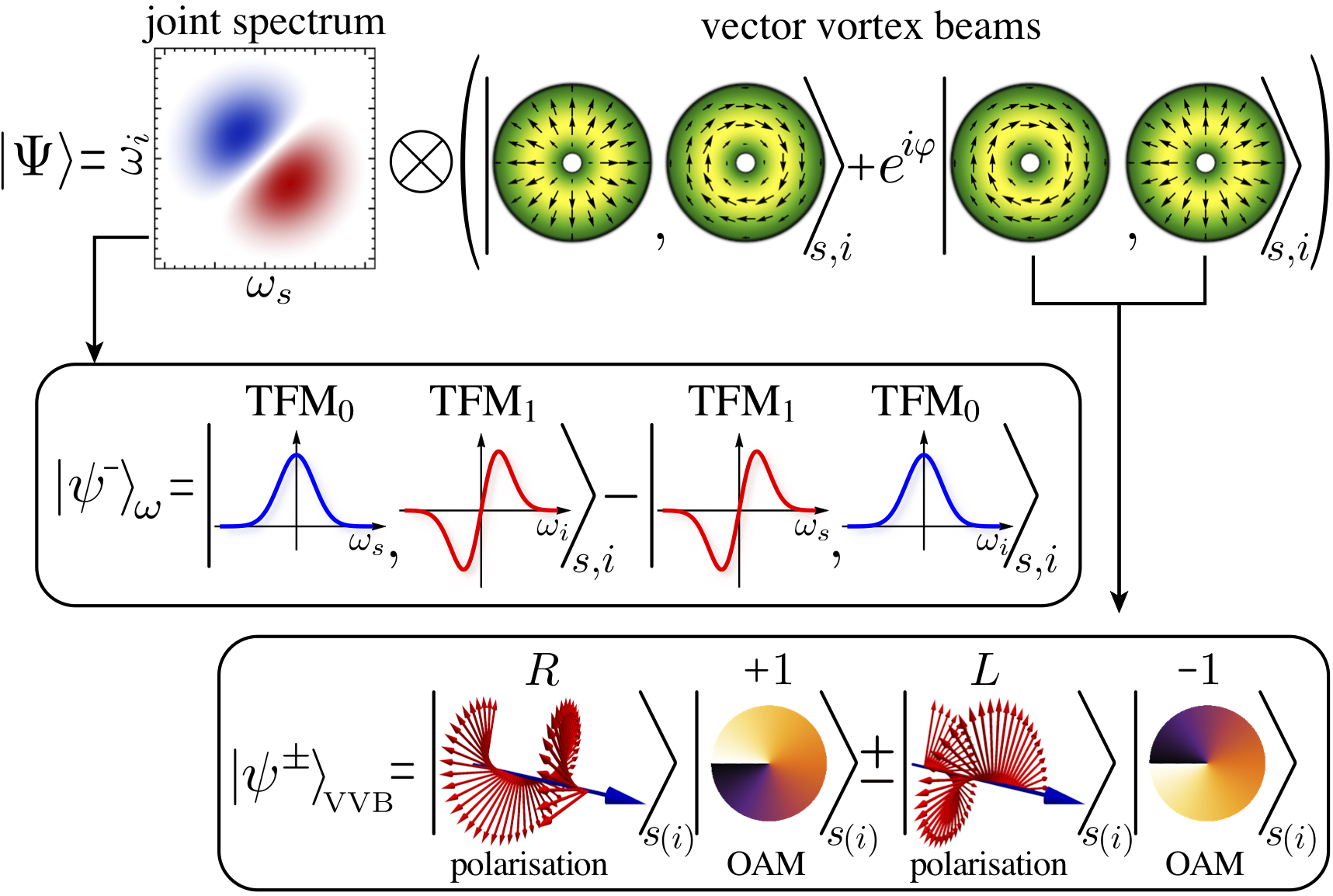}
\end{center}
\vspace{-1em}
\caption{\textbf{Sketch of the biphoton hyperentangled state.}
The overall quantum state $\ket{\Psi}$ (first row), exhibits hyperentanglement between spectrally and spatially structured light.
Signal and idler are encoded in the $\ket{\psi^-}_\omega$ state of the TFM basis, represented with the biphoton joint spectrum and the corresponding expansion in TFMs (first box), and in any Bell states of the VVB basis (we only display $\ket{\psi}$-type states for compactness).
Each photon is also in a single-particle entangled state between polarisation and OAM, giving rise to a VVB (second box). The plus (minus) sign corresponds to the radially (azimuthally) polarised beams, respectively.}
\label{fig:state}
\end{figure}

The generation of hyperentanglement between time-frequency modes and vector vortex beams implies the ability to independently create spectrally and spatially structured light.
TFM encoding can be achieved via nonlinearity engineering, a technique that tailors the phasematching function in parametric downconversion (PDC) processes by modifying the ferroelectric structure of periodically-poled nonlinear crystals.
Originally used for generating spectrally-pure heralded single photons~\cite{Branczyk:11,Graffitti_2017,Graffitti:18}, non-linearity engineering has since been applied to generate TFM entanglement with high fidelity~\cite{PhysRevLett.124.053603}.
Vector vortex beams on the other hand can be efficiently created by converting polarisation encoding into polarisation-OAM entanglement by means of birefringent liquid crystal devices known as q-plates~\cite{PhysRevLett.96.163905}. 
Our scheme combines these two techniques with an interferometric Sagnac scheme~\cite{Fedrizzi:07} for generating the highly-entangled state with the nontrivial structure sketched in Fig.~\ref{fig:state}.

\begin{figure}[b!]
\begin{center}
\includegraphics[width=0.90\columnwidth]{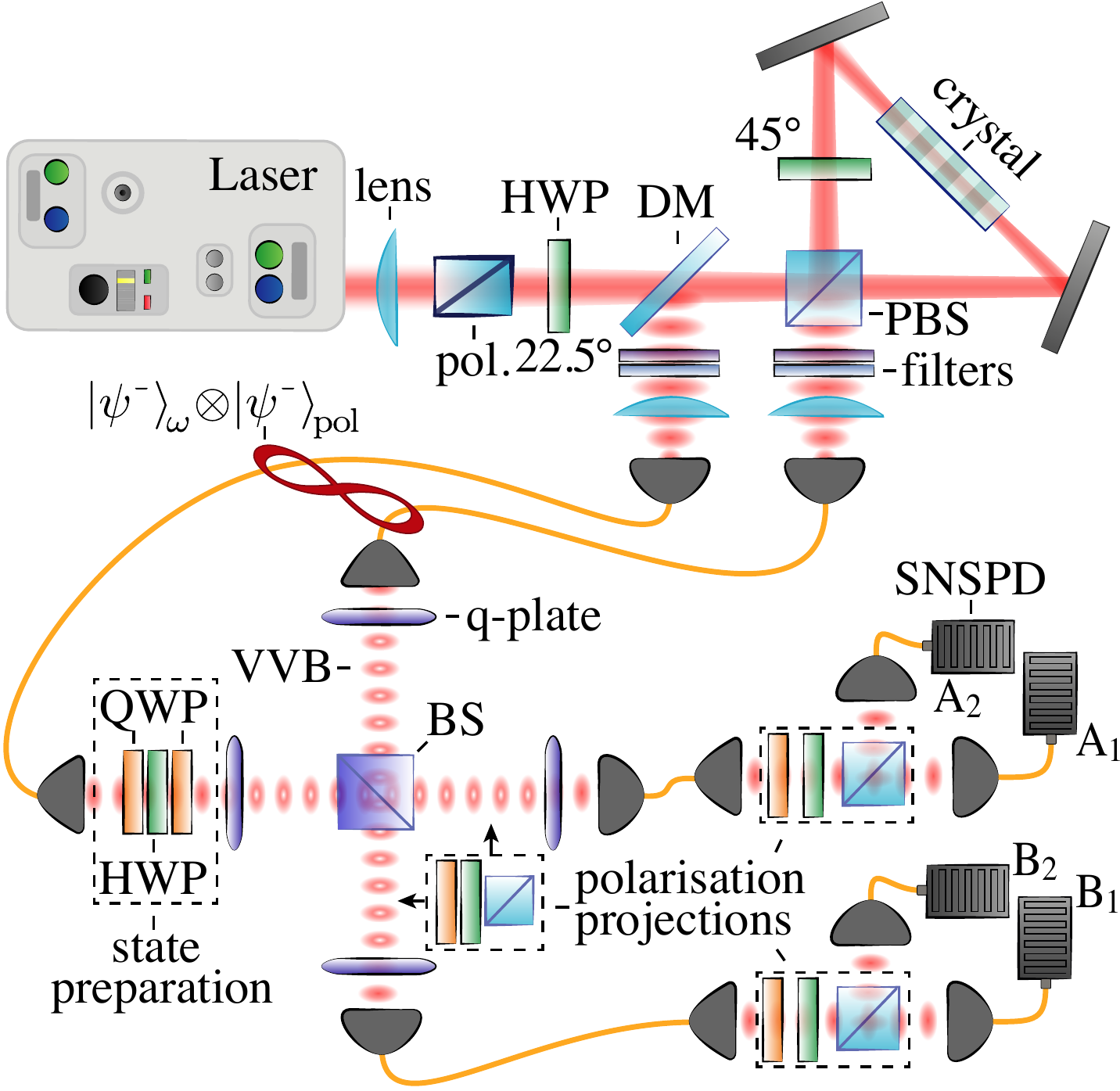}
\end{center}
\vspace{-1em}
\caption{\textbf{Experimental setup.}
Interferometric scheme to produce hyperentangled biphoton states in TFM and polarisation encoding, $\ket{ \psi^- }_\omega \otimes \ket{ \psi^-}_{\textrm{pol}}$ (top); Setup to produce  $\ket{\psi^-}_\omega \otimes \ket{\psi^\varphi}_\text{VVB}$ converting polarisation encoding into VVB encoding, and to measure the overall state via two-photon interference and tomographic reconstruction (bottom).
A set of QWP, HWP, QWP is used to prepare any maximally entangled state in the radial/azimuthal VVB basis. When fast axis of the QWPs is aligned, a rotation of the HWP corresponds to changing the phase factor of the Bell-like state.
The polarisation projection stages in the dashed boxes, each consisting of QWP, HWP, and polariser, are used to perform projective measurements on the polarisation of the photons for the sixteen-dimensional polarisation-OAM biphoton state reconstruction.
}
\label{fig:setup}
\end{figure}

We describe the experimental implementation in Fig.~\ref{fig:setup}.
A Ti-Sapphire laser produces a train of near transform-limited, $1.3$~ps pulses centred at $775$~nm with $80$~MHz repetition rate.
The laser is focused into a Sagnac interferometer for generating polarisation entanglement~\cite{Fedrizzi:07}, where a nonlinearity engineered crystal (details in Ref.~\cite{PhysRevLett.124.053603}) simultaneously enables the generation of TFM entanglement in a tailored PDC process
(see Sec.~1 of Supplemental Material for mathematical details on the state generation). 
This first section of the setup produces biphoton states that carry hyperentanglement between the maximally antisymmetric TFM Bell-state and a polarisation Bell-state. 
The PDC photons (signal ``s'' and idler ``i'') have a bandwidth of $\sim2.4$~nm (defined as the full-width at half maximum of the marginal photon's spectral intensity) and, after being separated at a polarising beamsplitter (PBS), are spectrally filtered with a long-pass filter (cut-off wavelength at $1400$~nm) and a ``loose'' bandpass filter
($10$~nm nominal bandwidth) before being coupled into single-mode fibres for spatial mode filtering.
A set of quarter-wave plate (QWP), half-wave plate (HWP) and QWP is used to prepare any maximally-entangled polarisation state via local operations.

Each photon is sent through a q-plate for converting polarisation encoding into VVB encoding, producing the target TFM-VVB hyperentangled state.
A q-plate with topological charge $q$ (with $q=0.5$ in our setup) implements the following transformation: $\alpha \ket{R ,0} + \beta \ket{L ,0} \to \alpha \ket{L ,-2q} + \beta \ket{R ,2q}$ where the first and second label correspond to polarisation and OAM value, respectively.
If the input polarisation is linear, a VVB in a linear superposition of the basis states $\ket{\hat{r}}$ (radially polarised) and  $\ket{\hat{\theta}}$ (azimuthally polarised) is produced in the process.
When each photon of polarisation-entangled biphoton state is sent through a q-plate, the overall system consists of two entangled vector vortex beams~\cite{PhysRevA.94.030304}.
At the output of the q-plates, the biphoton state shows a nontrivial entanglement structure, where three different forms of entanglement coexist in the same quantum state: hyperentanglement between TFMs and VVBs, which is in turn composed of single-particle (intrasystem) entanglement~\cite{Aiello_2015,PhysRevA.92.023833,PhysRevA.94.030304,PhysRevA.100.063842}---polarisation and OAM of each photon---and two distinct sets of intersystem entanglement---between the two VVBs and between the two TFMs,  as sketched in Fig.~\ref{fig:state}.
We note that this scheme allows one to generate states within a two-dimensional VVB subspace, additional HWPs after the q-plates enable the generation of any VVB state in the four-dimensional space~\cite{PhysRevA.94.030304}.

The analysis stage consists of two main steps.
First, we send the two photons on a BS to check for quantum interference depending on the symmetry of the full state~\cite{PhysRevLett.124.053603}.
After the BS, a set of two additional q-plates (with $q=0.5$) and polarisation optics (QWP, HWP and PBS for each photon) are used to perform tomographic projections in the VVB space.
The photons are finally detected with superconducting nanowire single-photon detectors (SNSPDs) with $80\%$ nominal quantum efficiency.
An additional tomographic projection set can be added before the measurement q-plates to perform a four-qubit tomography in the polarisation and OAM subspaces of the VVBs, simultaneously certifying the instrasystem entangled structure of each VVB and the intersystem entanglement between the two photons~\cite{PhysRevA.94.030304}, and verifying the GHZ-type structure of the state~\cite{Carvacho2017}.

\begin{figure}[t!]
\begin{center}
\includegraphics[width=0.90\columnwidth]{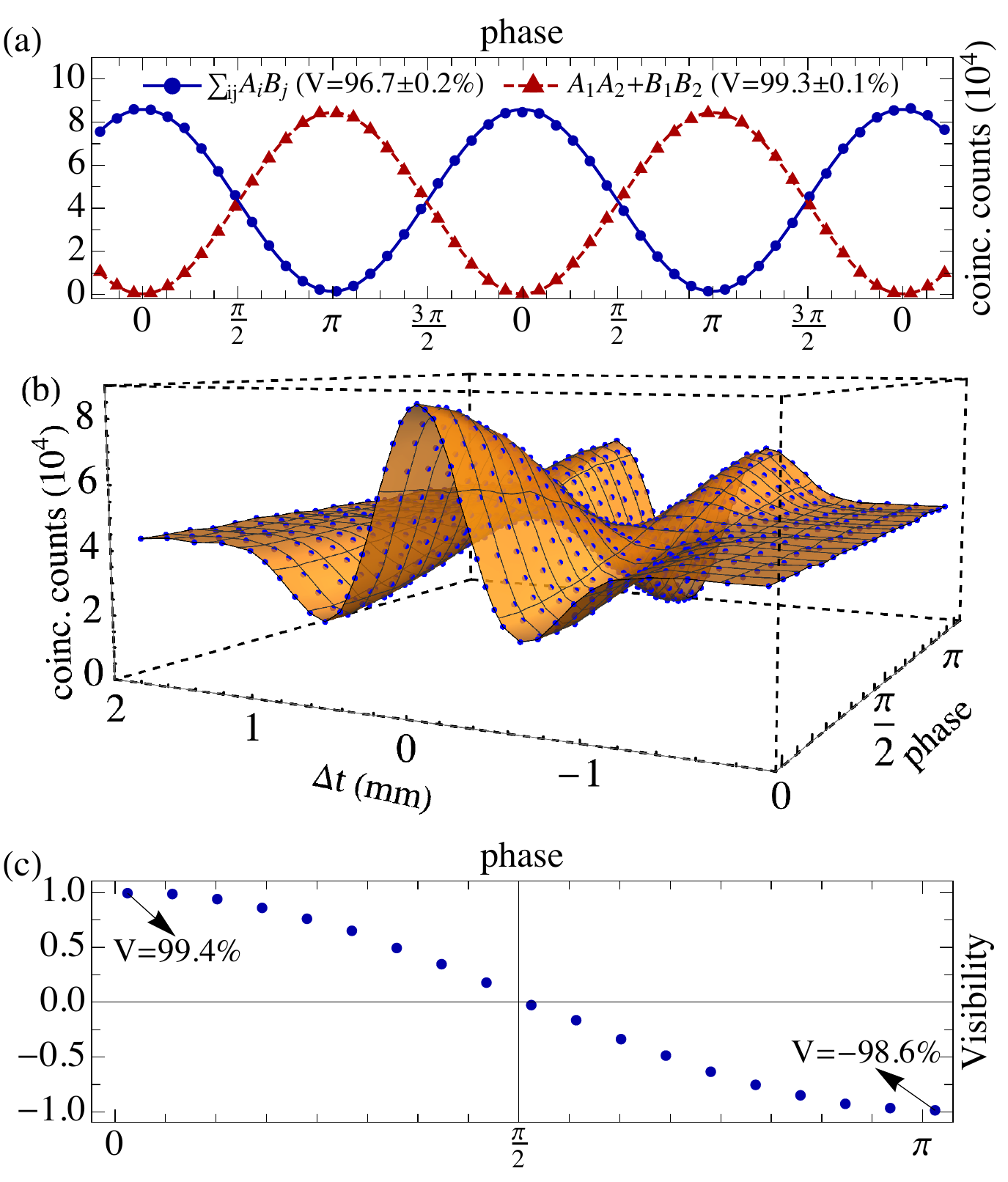}
\end{center}
\vspace{-1em}
\caption{\textbf{Two-photon interference results.}
\textbf{(a)} The interference fringes depend on the phase in the VVB-encoded part of the hyperentangled state:
A maximum in the coincident counts at the two outputs of the BS (labeled as $\sum_{ij}\!A_iB_j$, where $A_i$, $B_j$ are the detectors in Fig.~\ref{fig:setup}) corresponds to a minimum in the coincident counts at each BS output (labeled as $A_1A_2+B_1B_2$), and vice-versa.
Interference patterns collecting coincident counts at the two outputs of the BS \textbf{(b)} and corresponding visibilities \textbf{(c)} changing phase and relative arrival time of the photons at the BS. By controlling the phase of the $\ket{\psi}$-type we can 
move from almost perfect antibunching to bunching, i.e. from an overall antisymmetric state to a symmetric one.
}
\label{fig:results}
\end{figure}

We  produce states of the form $\ket{\psi^-}_\omega \otimes \ket{\psi^\varphi}_\text{VVB}$, where $\ket{\psi^-}_\omega = 1/\sqrt{2}\left( \ket{\hgzero, \hgone}_i - \ket{\hgone, \hgzero}_i \right)$ is the antisymmetric Bell-state in the TFM Bell-basis, and  $\ket{\psi^\varphi}_\text{VVB} = 1/\sqrt{2} \left( \ket{\hat{r},\hat{\theta}} + e^{i \varphi} \ket{\hat{\theta},\hat{r}} \right)$ is a $\psi$-type maximally entangled state in the VVB basis. 
By changing the HWP angle in the state preparation we can change the phase of the VVB part of the state and hence the symmetry of the overall wavefunction. The phase factors $e^{i 0}$ and $e^{i \pi}$ correspond to the $\ket{\psi^-}_\omega \otimes \ket{\psi^+}_\text{VVB}$ and $\ket{\psi^-}_\omega \otimes \ket{\psi^-}_\text{VVB}$ states, i.e. to a maximally antisymmetric and symmetric state, respectively.
This translates into different interference behaviour at the BS, where moving from an antisymmetric to a symmetric state corresponds to moving from photon antibunching to photon bunching, as we show in Fig.~\ref{fig:results}.

We monitor coincident counts between detectors $\left\{\text{A}_1,\text{B}_1\right\}$, $\left\{\text{A}_1,\text{B}_2\right\}$, $\left\{\text{A}_2,\text{B}_1\right\}$, $\left\{\text{A}_2,\text{B}_2\right\}$, 
corresponding to the photons exiting from both outputs of the BS, and between the detectors $\left\{\text{A}_1,\text{A}_2\right\}$, $\left\{\text{B}_1,\text{B}_2\right\}$, corresponding to the photons exiting the same outputs of the BS, to reconstruct the interference fringes as a function of the state's phase $\varphi$.
We show the results in Fig.~\ref{fig:results}~(a): the fringes corresponding to antibunching (blue dots) and to bunching (red triangles) are in antiphase, and have high visibilities ($96.7\pm0.2\%$ and $99.3\pm0.1\%$, respectively) certifying a high quality of the generated state.
By varying both the state's phase and the relative arrival time of signal and idler at the BS, we can reconstruct the full biphoton interference pattern for states with different amount of antisymmetry.
The 3D plot in Fig.~\ref{fig:results}(b) shows how the interference pattern changes from perfect antibunching (corresponding to $\varphi = 0$) to perfect bunching ($\varphi =  \pi$), in excellent agreement with the theoretical model we discuss in Sec.~2 of the Supplemental Material.
Finally, Fig.~\ref{fig:results}(c) shows the interference visibilities of each scan, where plus and minus $100\%$ correspond to perfect antibunching and bunching, respectively.

\begin{figure}[t!]
\begin{center}
\includegraphics[width=0.90\columnwidth]{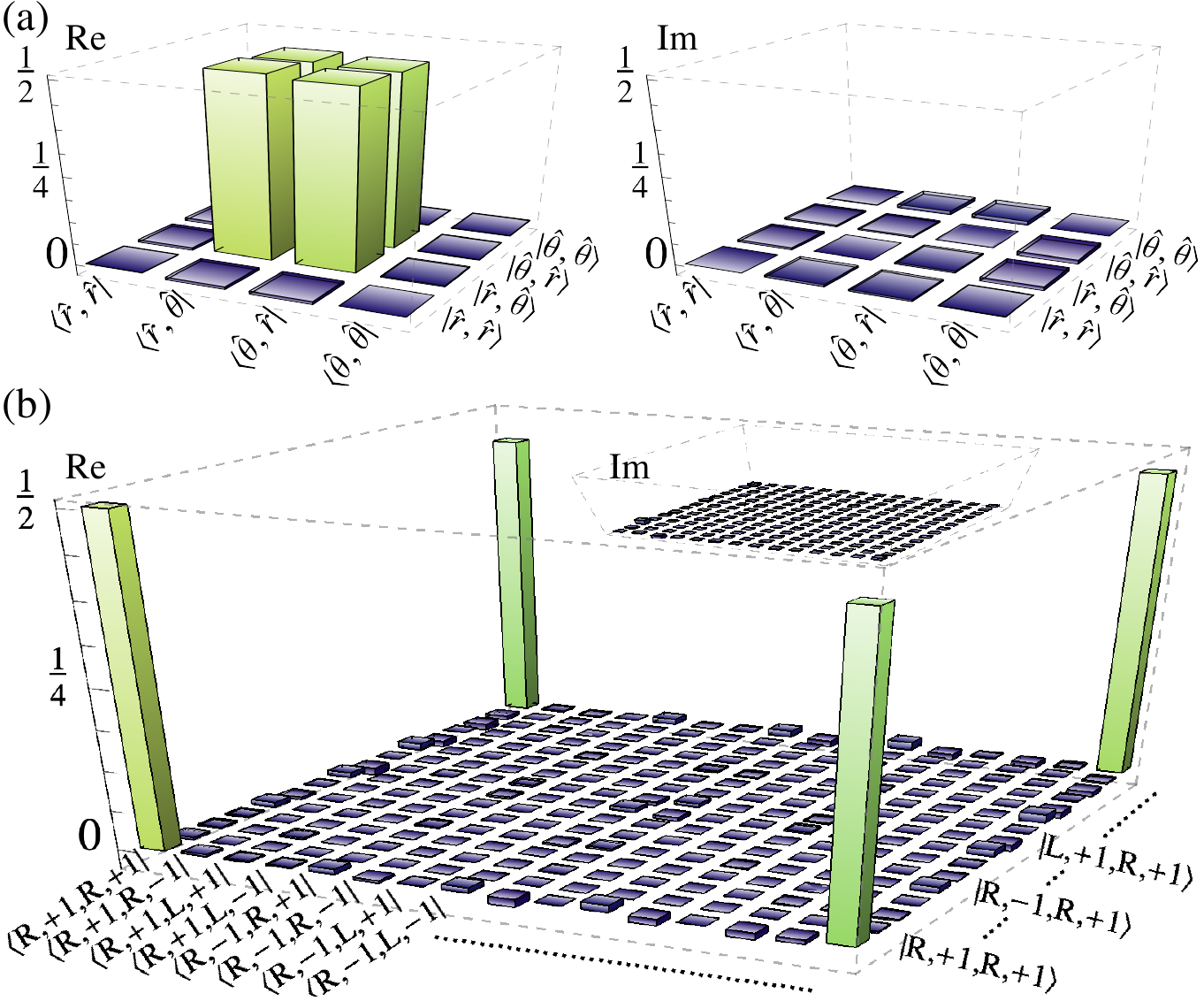}
\end{center}
\vspace{-1em}
\caption{\textbf{Tomography results.}
\textbf{(a)} Tomographic reconstruction of the biphoton $\ket{\psi^+}$ state in the VVB subspace $\left\{ \ket{\hat{r}}, \ket{\hat{\theta}} \right\}$.
\textbf{(b)} Tomographic reconstruction of the biphoton GHZ state encoded in polarisation and OAM.}
\label{fig:tomo}
\end{figure}

The two-photon interference allows us to measure the overall antisymmetry of the biphoton state, but it doesn't provide any information on its spatial structure.
The VVB state can instead be measured via quantum state tomography after the interference at the BS.
We prepare the symmetric state $\ket{\psi^-}_\omega \otimes \ket{\psi^+}_\text{VVB}$, which antibunches at the BS.
We then convert the VVB information into polarisation information, and we perform a overcomplete quantum state tomography of the state.
We measure a purity and fidelity of $(99.26\pm0.07)\%$ and $(99.57\pm 0.03)\%$, respectively, in the VVB subspace $\left\{ \ket{\hat{r}}, \ket{\hat{\theta}} \right\}$ (see Fig.~\ref{fig:tomo}(a)).
Introducing an additional tomographic projection before each measurement q-plate allows us to investigate the polarisation-OAM intrasystem entanglement and, at the same time, the two-photon intersystem entanglement~\cite{PhysRevA.94.030304}. With this scheme, we measure a two-photon, four-qubit purity of $(92.4\pm0.1)\%$ and a fidelity of $(95.0\pm0.1)\%$ with the GHZ state $1/\sqrt{2}(\ket{R,+1,R,+1} + \ket{L,-1,L,-1})$: we show the corresponding density matrix in Fig.~\ref{fig:tomo}(b).
The high interference visibility measured in both the bunching and antibunching configuration, combined with the high state quality obtained via tomographic reconstruction of the VVB-encoded state, testifies an unprecedented capability of generating and manipulating structured light encoded in three different degrees of freedom with very high efficiency and precision.

Many photonic quantum protocols rely on entanglement to carry out their tasks efficiently, therefore the capability of generating and manipulating complex entangled states of light with high precision is a fundamental requirement and a key challenge of quantum technologies.
Here, we tackled this problem by introducing and experimentally demonstrating a scheme for efficient generation of a complex entanglement structure between three DOFs of light: polarisation, orbital angular momentum and time-frequency modes. 
To our knowledge, neither TFM encoding nor VVB encoding have been combined with other degrees of freedom before, while our work introduces a simple, yet high-quality source of TFM-VVB hyperentanglement.
We expect our scheme will find applications in quantum communication schemes (where increased information capacity and noise resilience are obvious advantages) but also in different areas of quantum technologies, such as metrology or imaging, where both TFM and VVB encoding have already been independently used as resource~\cite{D'Ambrosio2013,PhysRevLett.121.090501}.

There are two main routes to go beyond the results of this work in the future.
On the one hand, it would be ideal to explore the intrinsic high-dimensionality of these DOFs, generating higher order OAM and TFM states to increase the information capacity of the biphoton state and investigate even more complex entanglement structures.
On the other hand, implementing quantum pulse gates or other schemes~\cite{Huang:13,PhysRevLett.120.213601,PhysRevX.5.041017,Reddy:18} for performing TFM-manipulation and measurements would allow one to fully exploit the potential of our technique.

\smallskip
\noindent \textbf{Acknowledgements}

\noindent This work was supported by the UK Engineering and Physical Sciences Research Council (Grant Nos. EP/N002962/1 and EP/T001011/1.). FG acknowledges studentship funding from EPSRC under Grant No. EP/L015110/1. 
Italian Ministry of Education, University and Research (MIUR) through the PRIN Project `INPhoPOL'.
European Union Horizon 2020 program, within the European Research Council (ERC) Grant No. 694683, PHOSPhOR.

\smallskip
\noindent \textbf{Additional information}

\noindent See Supplemental Material for supporting content.

%

%

\newpage
\clearpage
\newpage

\end{document}


\title{Hyperentanglement in structured quantum light - Supplemental Material}

\author{Francesco Graffitti}
\affiliation{Institute of Photonics and Quantum Sciences, School of Engineering and Physical Sciences, Heriot-Watt University, Edinburgh EH14 4AS, United Kingdom}

\author{Vincenzo D'Ambrosio}
\affiliation{Dipartimento di Fisica, Università di Napoli Federico II, Complesso Universitario di Monte S. Angelo, Via Cintia, 80126 Napoli, Italy}

\author{Massimiliano Proietti}
\affiliation{Institute of Photonics and Quantum Sciences, School of Engineering and Physical Sciences, Heriot-Watt University, Edinburgh EH14 4AS, United Kingdom}

\author{Joseph Ho}
\affiliation{Institute of Photonics and Quantum Sciences, School of Engineering and Physical Sciences, Heriot-Watt University, Edinburgh EH14 4AS, United Kingdom}

\author{Bruno Piccirillo}
\affiliation{Dipartimento di Fisica, Università di Napoli Federico II, Complesso Universitario di Monte S. Angelo, Via Cintia, 80126 Napoli, Italy}

\author{Corrado de Lisio}
\affiliation{Dipartimento di Fisica, Università di Napoli Federico II, Complesso Universitario di Monte S. Angelo, Via Cintia, 80126 Napoli, Italy}

\author{Lorenzo Marrucci}
\affiliation{Dipartimento di Fisica, Università di Napoli Federico II, Complesso Universitario di Monte S. Angelo, Via Cintia, 80126 Napoli, Italy}

\author{Alessandro Fedrizzi}
\affiliation{Institute of Photonics and Quantum Sciences, School of Engineering and Physical Sciences, Heriot-Watt University, Edinburgh EH14 4AS, United Kingdom}

\begin{abstract}
This document provides Supplemental Material to ``Hyperentanglement in structured quantum light''.
The document is structured as follows: In Section~1 we discuss the hyperentangled-state generation; In Section~2 we derive the two-photon interference pattern for the hyperentangled state.
\end{abstract}

\maketitle

\renewcommand{\theequation}{S\arabic{equation}}
\renewcommand{\thefigure}{S\arabic{figure}}
\renewcommand{\thetable}{S\arabic{table}}
\renewcommand{\thesubsection}{S\Roman{subsection}}
\setcounter{equation}{0}
\setcounter{figure}{0}
\setcounter{table}{0}
\setcounter{section}{0}

\section{I. Hyperentangled-state generation}
\subsection{Frequency entanglement}
\begin{figure}[htbp]
\begin{center}
\includegraphics[width=0.80\columnwidth]{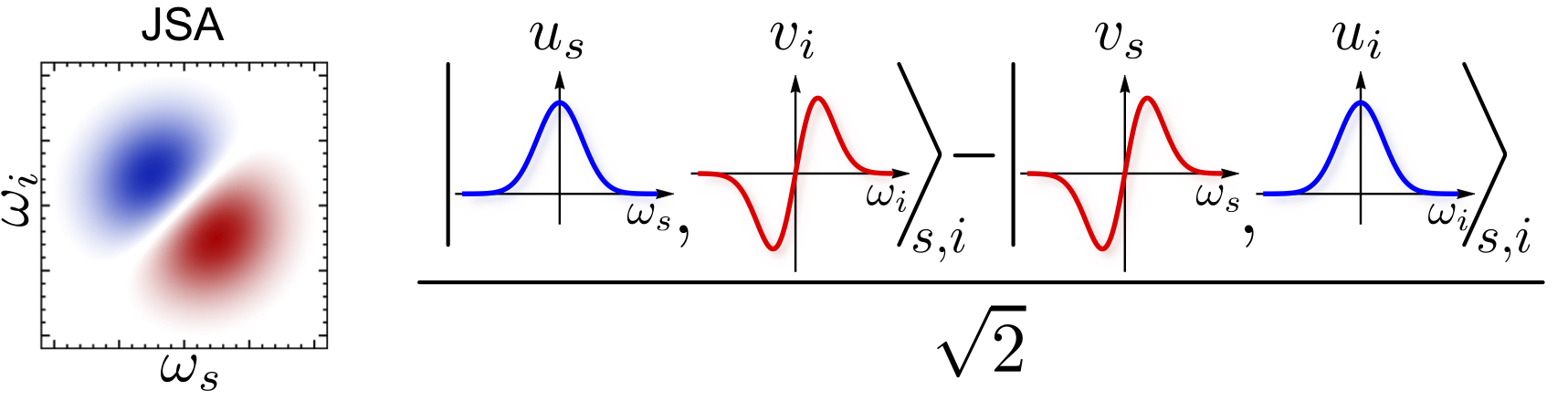}
\end{center}
\vspace{-1em}
\caption{\textbf{PDC biphoton state in frequency space.}
Joint spectral amplitude (left) and corresponding decomposed quantum state (right) encoded in two equally-weighted, orthonormal TFM.
}
\label{fig:jsa_modes}
\end{figure}

The PDC biphoton state, neglecting the multipair emission, reads:
\begin{equation}\begin{aligned}
\ket{\psi}&_{\omega}= 
\iint d\oos d\ooi f\ooa a_s^\dagger(\oos)a_i^\dagger(\ooi)\ket{0}_{s,i} \, ,
\label{eq:pdcbase}\end{aligned}\end{equation}
where $f\ooa$ is the joint spectral amplitude (JSA) and contains the spectral properties of the PDC biphoton state. 
The JSA is the product of the pump envelope function $\alpha \ooa$, describing the pump spectrum, and the phasematching function (PMF) $\phi\ooa$, depending on the nonlinear properties of the crystal that mediates the PDC process.
Here, we work in symmetric group-velocity matching condition (i.e. pump and PMF are perpendicular in the $\ooa$ plane~\cite{PhysRevA.98.053811}), and we consider a Gaussian-shaped pump pulse and a PMF equal to the first order Hermite-Gauss function multiplied by a Gaussian envelope
(deviations from the ideal case are discussed in details in Ref.~\cite{PhysRevLett.124.053603}). The corresponding JSA is shown in Fig.~\ref{fig:jsa_modes} (left).

This corresponds to the following PDC state:
\begin{equation}\begin{aligned}
\ket{\psi}_{\omega}=
\iint d\oos d\ooi \frac{2\left(\omega_s-\omega_i\right)}{\sqrt{\pi}\ \sigma^2} e^{-\frac{\omega_s^2+\omega_i^2}{\sigma^2}}  a_s^\dagger(\oos)a_i^\dagger(\ooi)\ket{0}_{s,i} \, .
\label{eq:pdcstate}\end{aligned}\end{equation}
We can perform the Schmidt decomposition on the state in Eq.~\eqref{eq:pdcstate} to obtain the TFMs composing the biphoton state, obtaining:
\begin{equation}\begin{aligned}
&\ket{u}_{\omega,j} 
\equiv \ket{\hgzero}_j = \left(\frac{2}{\pi\sigma^2}\right)^\frac{1}{4} \int d\omega_j  \ e^{-\frac{\omega_j^2}{\sigma^2}}\ a_j^\dagger(\omega_j)\ket{0}_{j}\\
&\ket{v}_{\omega,j} 
\equiv \ket{\hgone}_j = \left(\frac{2^5}{\pi\sigma^6}\right)^\frac{1}{4} \int d\omega_j  \ e^{-\frac{\omega_j^2}{\sigma^2}}\ \omega_j\  a_j^\dagger(\omega_j)\ket{0}_{j}\, ,
 \label{eq:schmidt}\end{aligned}\end{equation}
where $j$ can label either the signal $s$ or the idler $i$ photon, and there are only two non-null, balanced Schmidt coefficients, both equal to $\frac{1}{\sqrt{2}}$. 
Hence, the decomposed PDC state can be written as the maximally entangled singlet state:
\begin{equation}\begin{aligned}
\ket{\psi^-}_{\omega}
&= \frac{1}{\sqrt{2}}\left( \ket{u}_{\omega,s} \ket{v}_{\omega,i} - \ket{v}_{\omega,s}\ket{u}_{\omega,i} \right)
=\frac{1}{\sqrt{2}}\left( \ket{\hgzero}_s \ket{\hgone}_i - \ket{\hgone}_s\ket{\hgzero}_i \right)\, .
\label{eq:pdcsinglet}\end{aligned}\end{equation}
In Figure~\ref{fig:jsa_modes} (right) we show the two orthonormal TFMs in a balanced superposition.

\subsection{Polarisation-frequency hyperentanglement}

We now consider the case of a Sagnac interferometer source, commonly used for generating  polarisation entanglement in the two photon state~\cite{Fedrizzi:07}, where the standard ppKTP is replaced by the nonlinearity-engineered crystal in the Sagnac loop.
This enables the generation of an hyperentangled state in polarisation and frequency (as discussed in details in the next paragraph):
\begin{equation}\begin{aligned}
\ket{\Psi}_{\omega,{\scriptscriptstyle{\scriptscriptstyle\text{POL}}}}=
\frac{1}{\sqrt{2}}\left( \ket{\hgzero}_s \ket{\hgone}_i - \ket{\hgone}_s\ket{\hgzero}_i \right)\ \otimes \frac{1}{\sqrt{2}}\left( \ket{H}_s \ket{V}_i - \ket{V}_s\ket{H}_i \right)=\ket{\psi^-}_{\omega} \otimes \ket{\psi^-}_{{\scriptscriptstyle\text{POL}}}
\, ,
\label{eq:hyper}\end{aligned}\end{equation}
where the polarisation state $\ket{\psi^-_{\ {\scriptscriptstyle\text{POL}}}}_{s,i}$ can be easily manipulated via linear optical components (half- and quarter-wave plates) to generate any maximally entangled state.

Let's first consider both the clockwise and the anticlockwise cases where a pump photon is down-converted in two PDC photons having an antisymmetric joint spectral amplitude:
\begin{equation}
\begin{aligned}
&\ket{\psi_{\text{clockwise}}}_{\omega,{\scriptscriptstyle\text{POL}}} = \iint d\omega_s d\omega_i  f_{s,i}(\omega_s,\omega_i) c^\dagger_{s,H} c^\dagger_{i,V} \ket{0} \\
&\ket{\psi_{\text{anticlockwise}}}_{\omega,{\scriptscriptstyle\text{POL}}} = \iint d\omega_s d\omega_i  f_{s,i}(\omega_s,\omega_i) d^\dagger_{s,H} d^\dagger_{i,V} \ket{0} 
\, ,
\end{aligned} 
\end{equation}
where we omit the frequency dependence of the creation operators $c^\dagger$ (clockwise) and $d^\dagger$ (anticlockwise). 
The half-wave plate (HWP) and polarising beam splitter (PBS) introduce following transformations:
\begin{equation}
\begin{aligned}
&\text{HWP} : \begin{cases}
d^\dagger_{...,H} \to d^\dagger_{...,V} \\
d^\dagger_{...,V} \to d^\dagger_{...,H} 
\end{cases}\\
&\text{PBS} : \begin{cases}
c^\dagger_{...,H} \to a^\dagger_{...,H} \\
c^\dagger_{...,V} \to b^\dagger_{...,V} \\
d^\dagger_{...,H} \to b^\dagger_{...,H} \\
d^\dagger_{...,V} \to a^\dagger_{...,V} 
\end{cases}
\, ,
\end{aligned} 
\end{equation}
where $a^\dagger$,$b^\dagger$ corresponds to the two outputs of the PBS.
The state before entering the polarising beam splitter (PBS) reads:
\begin{equation}
\begin{aligned}
&\ket{\psi_{\text{clockwise}}}_{\omega,{\scriptscriptstyle\text{POL}}}^{\text{IN}} = \iint d\omega_s d\omega_i  f_{s,i}(\omega_s,\omega_i) c^\dagger_{s,H} c^\dagger_{i,V} \ket{0} \\
&\ket{\psi_{\text{anticlockwise}}}_{\omega,{\scriptscriptstyle\text{POL}}}^{\text{IN}} = \iint d\omega_s d\omega_i  f_{s,i}(\omega_s,\omega_i) d^\dagger_{s,V} d^\dagger_{i,H} \ket{0} 
\, ,
\end{aligned} 
\end{equation}
while at the output of the PBS the state is:
\begin{equation}
\begin{aligned}
&\ket{\psi_{\text{clockwise}}}_{\omega,{\scriptscriptstyle\text{POL}}}^{\text{OUT}} = \iint d\omega_s d\omega_i  f_{s,i}(\omega_s,\omega_i) a^\dagger_{s,H} b^\dagger_{i,V}  \ket{0} \\
&\ket{\psi_{\text{anticlockwise}}}_{\omega,{\scriptscriptstyle\text{POL}}}^{\text{OUT}} = \iint d\omega_s d\omega_i  f_{s,i}(\omega_s,\omega_i) a^\dagger_{s,V} b^\dagger_{i,H}  \ket{0} 
\, .
\end{aligned} 
\end{equation}
We note that in the Sagnac scheme the signal (idler) photons produced in both clockwise and anticlockwise paths exit the PBS from the same port, hence the labels $s$, $a$ (and $i$, $b$) are interchangeable.  
Being the clockwise and anticlockwise cases in a coherent superposition~\cite{Fedrizzi:07}, we can write the output state as:
\begin{equation}
\begin{aligned}
&\ket{\Psi}_{\omega,{\scriptscriptstyle\text{POL}}} = \frac{1}{\sqrt{2}} \left(\iint d\omega_s d\omega_i   f_{s,i}(\omega_s,\omega_i)  a^\dagger_{s,H} b^\dagger_{i,V}\ket{0}  + \iint d\omega_s d\omega_i f_{s,i}(\omega_s,\omega_i) a^\dagger_{s,V} b^\dagger_{i,H}  \ket{0}\right)\\
&\ket{\Psi}_{\omega,{\scriptscriptstyle\text{POL}}} = \frac{1}{\sqrt{2}} \left(\ket{\psi^-\ooa}_{s,i} \ket{H}_s\ket{V}_i + \ket{\psi^-\ooa}_{s,i} \ket{V}_s\ket{H}_i\right) \\
&\ket{\Psi}_{\omega,{\scriptscriptstyle\text{POL}}} = \ket{\psi^-\ooa}_{s,i} \otimes \frac{1}{\sqrt{2}} \left(\ket{H}_s\ket{V}_i + \ket{V}_s\ket{H}_i\right) = \ket{\psi^-\ooa}_{s,i}\ \otimes \ket{\psi^+_{\ {\scriptscriptstyle\text{POL}}}}_{s,i} 
\end{aligned} 
\end{equation}
where $\ket{\psi^+_{\ {\scriptscriptstyle\text{POL}}}(\omega_s,\omega_i)}_{s,i}$ is the symmetric polarisation-encoded Bell state.
From this state, one can prepare any state of the polarisation Bell basis by means of local operations:
\begin{equation}
\begin{aligned}
&\ket{\Psi^+ }_{\omega,{\scriptscriptstyle\text{POL}}} = \ket{\psi^-}_{\omega}\ \otimes \ket{\psi^+}_{{\scriptscriptstyle\text{POL}}}\\
&\ket{\Psi^- }_{\omega,{\scriptscriptstyle\text{POL}}} = \ket{\psi^-}_{\omega}\ \otimes \ket{\psi^-}_{{\scriptscriptstyle\text{POL}}} \\
&\ket{\Phi^+ }_{\omega,{\scriptscriptstyle\text{POL}}}= \ket{\psi^-}_{\omega}\ \otimes\ket{\phi^+}_{{\scriptscriptstyle\text{POL}}}\\
&\ket{\Phi^- }_{\omega,{\scriptscriptstyle\text{POL}}}= \ket{\psi^-}_{\omega}\ \otimes\ket{\phi^-}_{{\scriptscriptstyle\text{POL}}}\, .
\label{eq:hyper_bell}
\end{aligned} 
\end{equation}
$\Psi^+$, $\Phi^+$ and $\Phi^-$ are antisymmetric states, because they result from the product of an antisymmetric state (in TFM encoding) and a symmetric one (in polarisation encoding). $\Psi^-$ is the only symmetric state, being the product of two antisymmetric states.

\subsection{Polarisation to vector vortex beam conversion and final state}
When a polarisation-encoded photon is sent through a q-plate, the circular component of the polarisation is flipped and the photon acquires orbital angular momentum (OAM) according to the following transformation:
\begin{equation}\begin{aligned}
&\ket{L}\ket{0} \to \ket{R}\ket{+2q}\\
&\ket{R}\ket{0} \to \ket{L}\ket{-2q}
\, ,
 \label{eq:qplate}\end{aligned}\end{equation}
where the first and second ket state represent polarisation and OAM, respectively, and $q$ is the topological charge of the q-plate~\cite{PhysRevLett.96.163905}.
If the input polarisation of the photon is linear, the output photon is a vector vortex beam (VVB), i.e. polarisation and orbital angular momentum are entangled, giving rise to a non-uniform polarisation pattern in the transverse plane~\cite{Cardano:12}. In particular, by sending H and V polarised light through the q-plate one can generate the so called $\ket{\hat{r}}$ and $\ket{\hat{\theta}}$ states, respectively, defined as~\cite{PhysRevA.94.030304}:
\begin{equation}\begin{aligned}
&\ket{H} \to \ket{\hat{r}} = \frac{1}{\sqrt{2}}\left( \ket{R}\ket{+2q}+\ket{L}\ket{-2q}\right)\\
&\ket{V} \to \ket{\hat{\theta}}= \frac{1}{\sqrt{2}}\left( \ket{R}\ket{+2q}-\ket{L}\ket{-2q}\right)
\, .
 \label{eq:VVB}\end{aligned}\end{equation}
The other two state of the four-dimensional VVB basis, namely $\ket{\hat{\pi}^+}$ and $\ket{\hat{\pi}^-}$, can be generated with an additional HWP after the q-plate:
\begin{equation}\begin{aligned}
&\ket{\hat{\pi}^+} = \frac{1}{\sqrt{2}}\left( \ket{L}\ket{+2q}+\ket{R}\ket{-2q}\right)\\
&\ket{\hat{\pi}^-}= \frac{1}{\sqrt{2}}\left( \ket{L}\ket{+2q}-\ket{R}\ket{-2q}\right)
\, .
 \label{eq:piVVB}\end{aligned}\end{equation}

By sending a polarisation-TFM hyperentangled state (i.e. the one described in Eq.~\eqref{eq:hyper}) through the q-plates, the following state is produced:
\begin{equation}\begin{aligned}
\ket{\Psi}_{\omega,{\scriptscriptstyle\text{VVB}}}=\frac{1}{\sqrt{2}}\left( \ket{\hgzero}_s \ket{\hgone}_i - \ket{\hgone}_s\ket{\hgzero}_i \right)\ \otimes \frac{1}{\sqrt{2}}\left( \ket{\hat{r}}_s \ket{\hat{\theta}}_i - \ket{\hat{\theta}}_s\ket{\hat{r}}_i \right)
=\ket{\psi}_{\omega}\ \otimes\ \ket{\psi^-}_{{\scriptscriptstyle\text{VVB}}}
\, ,
\label{eq:hyperVVB}\end{aligned}\end{equation}
which is an hyperentangled between the TFM space and VVB space. In such system, intersystem hyperentanglement (between VVB and TFM encoded in signal and ilder) and intrasystem entanglement (between polarisation and OAM of each individual photon) coexist in the same biphoton state.

\section{2. Two-photon interference of the hyperentangled state}

In this section we discuss the two-photon interference for the polarisation-TFM hyperentangled state. We note that the same procedure can be used for the VVB-TFM hyperentanglement, where the polarisation indices $H$ and $V$ needs to be replaced by the VVB indices $\hat{r}$ and $\hat{\theta}$.

\begin{figure}[htbp]
\begin{center}
\includegraphics[width=0.70\columnwidth]{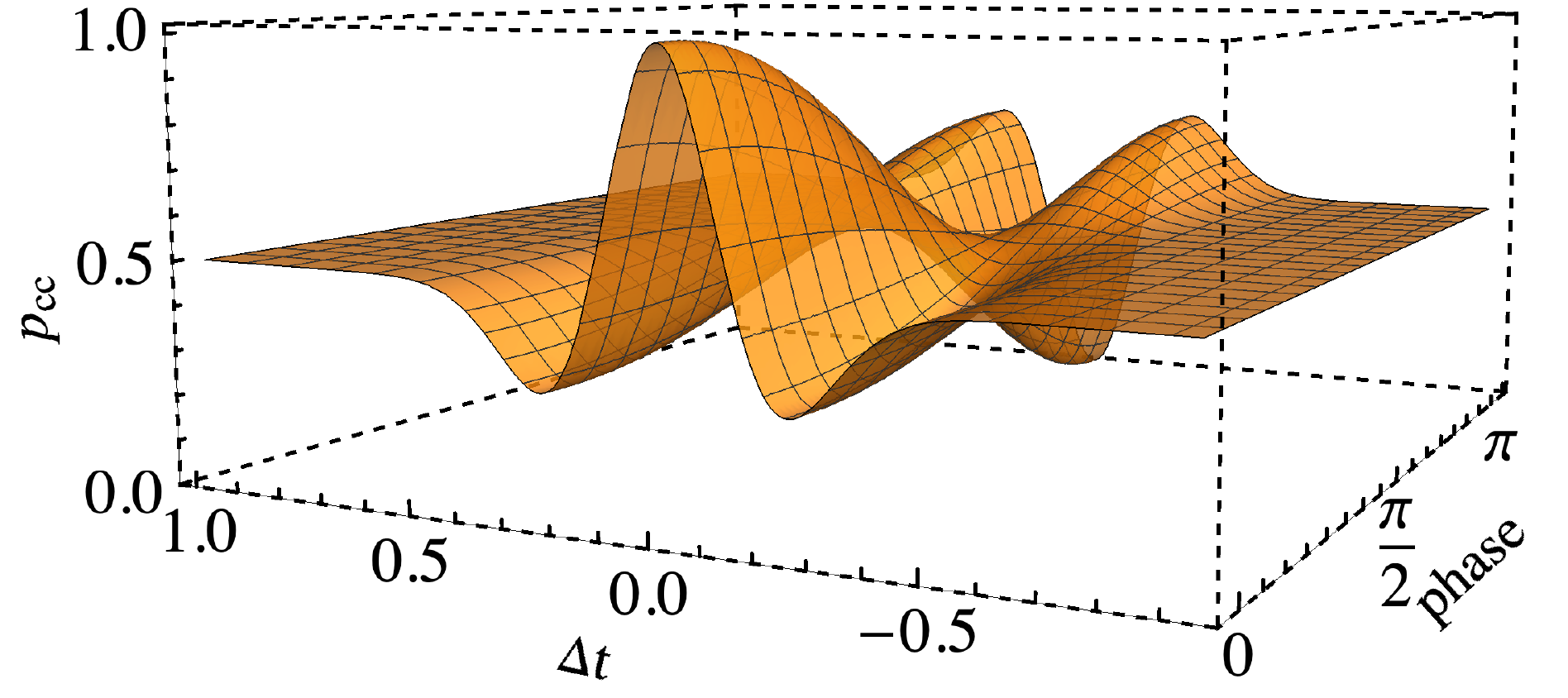}
\end{center}
\vspace{-1em}
\caption{\textbf{Two-photon interference pattern.}
Probability of having coincidences after the BS as a function of the state's phase, varying the relative arriving time of signal and idler.
}
\label{fig:hom3d}
\end{figure}

Let's consider the hyperentangled state:
\begin{equation}
\begin{aligned}
\ket{\Psi}_{\omega,{\scriptscriptstyle\text{POL}}} = \ket{\psi^-}_{\omega} \otimes \frac{1}{\sqrt{2}} \left( \ket{H}_s \ket{V}_i + e^{i \varphi} \ket{V}_s\ket{H}_i \right) \, :
\label{eq:HOMinput_state}
\end{aligned} 
\end{equation}
depending on the value of the phase $\varphi$, the overall quantum state can be maximally symmetric ($\varphi = \pi$), maximally antisymmetric ($\varphi = 0$), or a combination of the two cases (for any other value of $\varphi$).
The signal and idler photon enter the ports $a$ and $b$ of a beam splitter (BS), which introduce the following transformations:
\begin{equation}
\begin{aligned}
\text{BS} : \begin{cases}
a^\dagger_{p}(\omega_s) \to \frac{1}{\sqrt{2}}\left( ia^\dagger_{p}(\omega_s)  + b^\dagger_{p}(\omega_s) \right) \\
b^\dagger_{p} (\omega_i) \to  \frac{1}{\sqrt{2}}\left(a^\dagger_{p}(\omega_i) + i b^\dagger_{p}(\omega_i)\right)
\end{cases}
\end{aligned} 
\end{equation}
where $p$ is the polarisation index.
The state after the BS reads:
\begin{equation}
\begin{gathered}
\ket{\Psi }_{\omega,{\scriptscriptstyle\text{POL}}} = \frac{1}{\sqrt{2}}
\iint d\omega_s d\omega_i f_{s,i}(\omega_s,\omega_i) 
e^{-i\omega_i \tau}
\left(  a^\dagger_{H}(\omega_s)b^\dagger_{V}(\omega_i) + e^{i \varphi} a^\dagger_{V}(\omega_s)b^\dagger_{H}(\omega_i)\right)
\ket{0}\\
\Downarrow \text{BS} \\
\ket{\Psi }_{\omega,{\scriptscriptstyle\text{POL}}} = \frac{1}{2\sqrt{2}}
\iint d\omega_s d\omega_i f_{s,i}(\omega_s,\omega_i) 
e^{-i\omega_i \tau}\\
 \left(  \left( ia^\dagger_{H}(\omega_s)  + b^\dagger_{H}(\omega_s) \right) \left(a^\dagger_{V}(\omega_i) + i b^\dagger_{V}(\omega_i)\right) + e^{i \varphi} \left( ia^\dagger_{V}(\omega_s)  + b^\dagger_{V}(\omega_s) \right)\left(a^\dagger_{H}(\omega_i) + i b^\dagger_{H}(\omega_i)\right)        \right)
\ket{0}
\, ;
\end{gathered} 
\label{stateBS}
\end{equation}
where we added a delay $\tau$ on the arrival time of the idler photon respect to the signal, and we decomposed the state in Eq.~\eqref{eq:HOMinput_state} in its spectral and polarisation components.
We can expand \eqref{stateBS} to:
\begin{equation}
\begin{aligned}
\ket{\Psi }_{\omega,{\scriptscriptstyle\text{POL}}} = \frac{1}{2\sqrt{2}}
&\iint d\omega_s d\omega_i f_{s,i}(\omega_s,\omega_i) 
e^{-i\omega_i \tau} \\
&\times \left( i a^\dagger_{H}(\omega_s) a^\dagger_{V}(\omega_i) -  a^\dagger_{H}(\omega_s) b^\dagger_{V}(\omega_i)  +  b^\dagger_{H}(\omega_s) a^\dagger_{V}(\omega_i) + i b^\dagger_{H}(\omega_s) b^\dagger_{V}(\omega_i) +\right.\\
&+ \left. e^{i \varphi} ia^\dagger_{V}(\omega_s) a^\dagger_{H}(\omega_i) - e^{i \varphi} a^\dagger_{V}(\omega_s) b^\dagger_{H}(\omega_i) + e^{i \varphi} b^\dagger_{V}(\omega_s)a^\dagger_{H}(\omega_i) + e^{i \varphi} i b^\dagger_{V}(\omega_s)b^\dagger_{H}(\omega_i)\right)
\ket{0} \,.
\end{aligned}
\end{equation}
We now project on the all possible coincidence cases:
\begin{equation}
\begin{aligned}
\hat{P}_{s,i}=\iint d\omega_a \omega_b \left( a^\dagger_{H}(\omega_a) b^\dagger_{V}(\omega_b) \ket{0}\bra{0}  a_{H}(\omega_a) b_{V}(\omega_b) + a^\dagger_{V}(\omega_a) b^\dagger_{H}(\omega_b) \ket{0}\bra{0}  a_{V}(\omega_a) b_{H}(\omega_b)    \right) \, ,
\end{aligned}
\end{equation}
where we neglect the cases where the operators have the same label ($(H,H)$ and $(V,V)$) because they are not present in the state.
\begin{equation}
\begin{aligned}
p_{cc} (\tau) = &\text{Tr}\left[  \ket{\Psi}\bra{\Psi }_{\omega,{\scriptscriptstyle\text{POL}}} \hat{P}_{s,i}  \right] =
\bra{\Psi}   \hat{P}_{s,i}   \ket{\Psi }_{\omega,{\scriptscriptstyle\text{POL}}} \\
&=\frac{1}{8} \iint d\omega_s d\omega_i f^*_{s,i}(\omega_s,\omega_i) e^{i\omega_i \tau} 
\bra{0}  \left( -i a_{H}(\omega_s) a_{V}(\omega_i) -  a_{H}(\omega_s) b_{V}(\omega_i)  +  b_{H}(\omega_s) a_{V}(\omega_i) - i b_{H}(\omega_s) b_{V}(\omega_i) +\right.\\
&\left.-e^{-i \varphi} ia_{V}(\omega_s) a_{H}(\omega_i) -e^{-i \varphi} a_{V}(\omega_s) b_{H}(\omega_i) +e^{-i \varphi}b_{V}(\omega_s)a_{H}(\omega_i) -e^{-i \varphi}  i b_{V}(\omega_s)b_{H}(\omega_i)\right)\\
&\times \iint d\omega_a \omega_b \left( a^\dagger_{H}(\omega_a) b^\dagger_{V}(\omega_b) \ket{0}\bra{0}  a_{H}(\omega_a) b_{V}(\omega_b) + a^\dagger_{V}(\omega_a) b^\dagger_{H}(\omega_b) \ket{0}\bra{0}  a_{V}(\omega_a) b_{H}(\omega_b)    \right) \\
&\times \iint d\omega'_s d\omega'_i f_{s,i}(\omega'_s,\omega'_i) 
e^{-i\omega'_i \tau} \\
&\left( i a^\dagger_{H}(\omega'_s) a^\dagger_{V}(\omega'_i) -  a^\dagger_{H}(\omega'_s) b^\dagger_{V}(\omega'_i)  +  b^\dagger_{H}(\omega'_s) a^\dagger_{V}(\omega'_i) + i b^\dagger_{H}(\omega'_s) b^\dagger_{V}(\omega'_i) +\right.\\
&\left.+ e^{i \varphi} ia^\dagger_{V}(\omega'_s) a^\dagger_{H}(\omega'_i) - e^{i \varphi} a^\dagger_{V}(\omega'_s) b^\dagger_{H}(\omega'_i) +e^{i \varphi}b^\dagger_{V}(\omega'_s)a^\dagger_{H}(\omega'_i) +e^{i \varphi}  i b^\dagger_{V}(\omega'_s)b^\dagger_{H}(\omega'_i)\right)
\ket{0} \, .
\end{aligned}
\end{equation} 
We can rearrange the terms obtaining:
\begin{equation}
\begin{aligned}
p_{cc} (\tau) =& \frac{1}{8} \iint d\omega_s d\omega_i d\omega'_s d\omega'_i d\omega_a \omega_b \\
&\times f^*_{s,i}(\omega_s,\omega_i) f_{s,i}(\omega'_s,\omega'_i)   e^{i(\omega_i - \omega'_i) \tau} \\
\times &\left(-\bra{0}   a_{H}(\omega_s) b_{V}(\omega_i)  a^\dagger_{H}(\omega_a) b^\dagger_{V}(\omega_b) \ket{0}\right.\\
&+\bra{0} b_{H}(\omega_s) a_{V}(\omega_i) a^\dagger_{V}(\omega_a) b^\dagger_{H}(\omega_b) \ket{0}\\
&-e^{-i \varphi}\bra{0}a_{V}(\omega_s) b_{H}(\omega_i)  a^\dagger_{V}(\omega_a) b^\dagger_{H}(\omega_b) \ket{0}\\
&\left.+e^{-i \varphi}\bra{0} b_{V}(\omega_s)a_{H}(\omega_i)  a^\dagger_{H}(\omega_a) b^\dagger_{V}(\omega_b) \ket{0}\right)\\
\end{aligned}
\quad \times \quad 
\begin{aligned}
&\\&\\&\\
&\left(-\bra{0} a_{H}(\omega_a) b_{V}(\omega_b) a^\dagger_{H}(\omega'_s) b^\dagger_{V}(\omega'_i)  \ket{0}\right.\\
&+\bra{0} a_{V}(\omega_a) b_{H}(\omega_b) b^\dagger_{H}(\omega'_s) a^\dagger_{V}(\omega'_i) \ket{0}\\
&-e^{i \varphi}\bra{0}  a_{V}(\omega_a) b_{H}(\omega_b) a^\dagger_{V}(\omega'_s) b^\dagger_{H}(\omega'_i) \ket{0}\\
&\left. +e^{i \varphi}\bra{0}  a_{H}(\omega_a) b_{V}(\omega_b) b^\dagger_{V}(\omega'_s)a^\dagger_{H}(\omega'_i)  \ket{0}\right)
\end{aligned}
\end{equation} 
where we have already neglected all the terms where the polarisation of the creation and annihilation operators don't match, and where both the creation operator/annihilation operators are in one mode (as we are considering coincidences between the two modes).
\begin{equation}
\begin{aligned}
&p_{cc} (\tau) = \frac{1}{8} \iint d\omega_s d\omega_i d\omega'_s d\omega'_i d\omega_a \omega_b 
f^*_{s,i}(\omega_s,\omega_i) f_{s,i}(\omega'_s,\omega'_i)   e^{i(\omega_i - \omega'_i) \tau} \\
&\times \left(
-\delta_{H}(\omega_s-\omega_a)\delta_{V}(\omega_i-\omega_b)
+\delta_{H}(\omega_s-\omega_b) \delta_{V}(\omega_i-\omega_a)-e^{-i \varphi}
\delta_{V}(\omega_s-\omega_a) \delta_{H}(\omega_i-\omega_b)+e^{-i \varphi}
\delta_{V}(\omega_s-\omega_b) \delta_{H}(\omega_i-\omega_a)\right)\\
&\times \left(
-\delta_{H}(\omega'_s-\omega_a)\delta_{V}(\omega'_i-\omega_b)
+\delta_{H}(\omega'_s-\omega_b) \delta_{V}(\omega'_i-\omega_a)-e^{i \varphi}
\delta_{V}(\omega'_s-\omega_a) \delta_{H}(\omega'_i-\omega_b)+e^{i \varphi}
\delta_{V}(\omega'_s-\omega_b) \delta_{H}(\omega'_i-\omega_a)\right)
\end{aligned}
\end{equation} 
Integrating over $\omega_a$ and $\omega_b$ we get:
\begin{equation}
\begin{aligned}
&p_{cc} (\tau) = \frac{1}{8} \iint d\omega_s d\omega_i d\omega'_s d\omega'_i
f^*_{s,i}(\omega_s,\omega_i) f_{s,i}(\omega'_s,\omega'_i)   e^{i(\omega_i - \omega'_i) \tau} \\
&\times \left(\delta_{H}(\omega_i-\omega'_i) \delta_{V}(\omega_s-\omega'_s)  -e^{i \varphi}
\delta_{H}(\omega_s-\omega'_i) \delta_{V}(\omega_i-\omega'_s)  +
\delta_{H}(\omega_s-\omega'_s) \delta_{V}(\omega_i-\omega'_i) - e^{i \varphi}
\delta_{H}(\omega_s-\omega'_i) \delta_{V}(\omega_i-\omega'_s)\right. +\\
&-e^{-i \varphi} \delta_{V}(\omega_s-\omega'_i) \delta_{H}(\omega_i-\omega'_s) + 
\delta_{V}(\omega_s-\omega'_s) \delta_{H}(\omega_i-\omega'_i)-e^{-i \varphi}
\delta_{V}(\omega_s-\omega'_i) \delta_{H}(\omega_i-\omega'_s)+
\delta_{V}(\omega_s-\omega'_s) \delta_{H}(\omega_i-\omega'_i)
\left.\right)
\end{aligned}
\end{equation} 
where we have neglected all the terms where the polarisation of the delta functions doesn't match.
Let's now integrate over $\omega'_s$ and $\omega'_i$ (and we can also get rid of the polarisation labels as are now irrelevant):
\begin{equation}
\begin{aligned}
p_{cc} (\tau) =& \frac{1}{8} \iint d\omega_s d\omega_i
\left(4 \abs{f_{s,i}(\omega_s,\omega_i)}^2 - 2 f_{s,i}^*(\omega_s,\omega_i)f_{s,i}(\omega_i,\omega_s) e^{i(\omega_i - \omega_s) \tau} e^{i\varphi} - 2 f_{s,i}^*(\omega_s,\omega_i)f_{s,i}(\omega_i,\omega_s) e^{i(\omega_i - \omega_s) \tau} e^{-i\varphi} \right)\\
=&\frac{1}{2} \iint d\omega_s d\omega_i
\left( \abs{f_{s,i}(\omega_s,\omega_i)}^2 -\cos{\left(\varphi\right)}  f_{s,i}^*(\omega_s,\omega_i)f_{s,i}(\omega_i,\omega_s) e^{i(\omega_i - \omega_s) \tau}  \right) \\
=&\frac{1}{2} \iint d\omega_s d\omega_i
\left( 1 - \cos{\left(\varphi\right)} f_{s,i}^*(\omega_s,\omega_i)f_{s,i}(\omega_i,\omega_s) e^{i(\omega_i - \omega_s) \tau}  \right) \, .
\end{aligned}
\end{equation}
We can finally write the JSA in terms of the Schmidt modes:
\begin{equation}
\begin{aligned}
p_{cc} (\tau) =& \frac{1}{2} \iint d\omega_s d\omega_i
\left( 1 - \cos{\left(\varphi\right)} \left( \sum_k d_k u^*_k (\omega_s) v^*_k (\omega_i) \right) \left(\sum_{k'} d_{k'} u_{k'} (\omega_i) v_{k'} (\omega_s)\right)  e^{i(\omega_i - \omega_s) \tau} \right)  \\
=& \frac{1}{2}   
\left( 1 - \cos{\left(\varphi\right)}  \left(  \sum_{k,k'} d_k d_{k'} \int d\omega_s u^*_k (\omega_s) v_{k'} (\omega_s)  e^{ - i \omega_s \tau}  \int d\omega_i  v^*_k (\omega_i)  u_{k'} (\omega_i)  e^{i\omega_i  \tau}  \right)\right)   \, .
\end{aligned}
\label{antisymm_HOM_decomposed}
\end{equation} 
Replacing $u$ and $v$ with the corresponding Schmidt modes in Eq.~\eqref{eq:schmidt} we obtain the following interference pattern:
\begin{equation}
\begin{aligned}
p_{cc}(\tau)=\frac{1}{2}-\frac{1}{4} e^{-\frac{1}{4}\sigma^2\tau^2}(\sigma^2\tau^2-2) \cos{\left(\varphi\right)}  \, ,
\end{aligned}
\label{hyper_HOM}
\end{equation}
that we plot in Fig.~\ref{fig:hom3d}.
$\varphi=0$ corresponds to perfect antibunching, as expected by a maximally antisymmetric state, while for $\varphi=\pi$ leads to perfect bunching, as the state is symmetric.

